\documentclass[preprint2]{aastex}

\usepackage{amsmath, amssymb, amsfonts}
\usepackage{natbib}
\usepackage{morefloats}
\usepackage[usenames, dvipsnames]{color}
\usepackage[bookmarks=true]{hyperref}
\hypersetup{
colorlinks=true,
linkcolor=magenta,
pdfborder={1 0 0},
citecolor=blue,
filecolor=blue,
urlcolor=blue,
bookmarksnumbered=true
}
\hypersetup{colorlinks=true,linkcolor=blue,citecolor=blue,filecolor=blue,urlcolor=blue}

\newcommand \beq {\begin{equation}}
\newcommand \enq {\end{equation}}
\newcommand \nush {\nu_{\rm sh}}
\newcommand \Csh  { C_{\rm sh}}
\newcommand \Qsh  { Q_{\rm sh}}
\newcommand \qsh  { q_{\rm sh}}

\newcommand \omgb { \Omega_{\rm bin}}
\newcommand \omgbi  {\Omega_{\rm bin}^{-1}}

\newcommand \vz    {v_{\rm {z}}}

\newcommand \Bz  {B_{\rm {z}}}

\slugcomment{}

\shorttitle{How bright are the gaps in circumbinary disk systems?}
\shortauthors{Shi, \& Krolik}

\begin{document}

%% LaTeX will automatically break titles if they run longer than
%% one line. However, you may use \\ to force a line break if
%% you desire.

\title{How bright are the gaps in circumbinary disk systems?}

%% Use \author, \affil, and the \and command to format
%% author and affiliation information.
%% Note that \email has replaced the old \authoremail command
%% from AASTeX v4.0. You can use \email to mark an email address
%% anywhere in the paper, not just in the front matter.
%% As in the title, use \\ to force line breaks.

\email{jmshi@astro.princeton.edu}
\author{Ji-Ming Shi\altaffilmark{1} and Julian H. Krolik\altaffilmark{2}}
\altaffiltext{1}{Department of Astrophysical Sciences, Princeton University, 4 Ivy Lane,
    Princeton, NJ 08544}
%\altaffiltext{2}{Department of Astronomy, UC Berkeley, Hearst Field Annex B-20,
%    Berkeley, CA 94720-3411}
%\altaffiltext{2}{Center for Integrative Planetary Science, UC Berkeley, Hearst Field Annex B-20,
%    Berkeley, CA 94720-3411}
\altaffiltext{2}{Department of Physics and Astronomy, Johns Hopkins
University, Baltimore, MD 21218}

\begin{abstract}
When a circumbinary disk surrounds a binary whose secondary's mass is at least $\sim 10^{-2}\times$ the primary's mass, a nearly empty cavity with radius a few times the binary separation is carved out of the disk.  Narrow streams of material pass from the inner edge of the circumbinary disk into the domain of the binary itself, where they eventually join onto the small disks orbiting the members of the binary.  Using data from 3-d MHD simulations of this process, we determine the luminosity of these streams; it is mostly due to weak laminar shocks, and is in general only a few percent of the luminosity of adjacent regions of either the circumbinary disk or the ``mini-disks". This luminosity therefore hardly affects the deficit in the thermal continuum predicted on the basis of a perfectly dark gap region.
\end{abstract}

%% Keywords should appear after the \end{abstract} command. The uncommented
%% example has been keyed in ApJ style. See the instructions to authors
%% for the journal to which you are submitting your paper to determine
%% what keyword punctuation is appropriate.


\section{Introduction}\label{sec:intro}

%I. Ubiquity of circumbinary disks with gaps
%   A. gaps occur whenever the binary mass ratio isn't too far from 1
%   B. seen in binary star-formation
%   C. seen in supermassive binary black holes
%   D. only matter in gaps: streams from inner edge of circumbinary disk to outer edges
%      of mini-disks

Circumbinary disks are ubiquitous astronomical objects. They are often observed around young binary stars \citep[e.g.,][]{Dutreyetal1994,AW2005} and may have already been detected in disk-planet systems \citep{Reggianietal2014,Billeretal2014,Sallumetal2015}. They may also be present around supermassive binary black holes (SMBBHs) as galaxies merge \citep{BBR1980,Ivanovetal1999,MeMi2005}. When the mass ratio is not too far from unity, the binary clears out a low density gap in the disk center, crossed by one or two narrow streams.  Emanating from the inner edge of the circumbinary disk, some of the matter in these streams reaches ``mini-disks" attached to the members of the binary, while some of the matter suffers strong torques and swings back out to the circumbinary disk.  \citep{AL1994,GK2002,MM2008,Cuadraetal2009,Hanawaetal,deValBorroetal2011, Roedigetal2012,Shietal2012,DOrazioetal2013,Farrisetal2014,ShiKrolik2015, DOrazio2016}.

%II. Natural question: does this distinctive region produce a distinctive light signal?
%   A. "notch" in thermal disk spectrum because streams are cold?
%   B. high-temperature feature from stream--mini-disk impact?
%   C. shock heating of the inflow and returning streams ?

A natural question to ask is therefore: does this distinctive region produce a characteristic light signal? There is some controversy about the answer to this question.  A number of papers have argued that the gap region of a circum-SMBBH disk, including the streams, should be relatively dim, and would therefore produce a dip in the thermal disk spectrum over a specific range of wavelengths dependent upon the parameters of the system \citep{RoedigSesana2012,Tanakaetal2012,GM2012,Kocsisetal2012,TH2013CQGra, Roedigetal2014}. Others have pointed to events on either the inside or the outside of the gap as creating distinct radiative features.  It has been suggested, for example, that because the accretion rate across the gap is in general modulated strongly at a frequency comparable to the orbital frequency, so, too, should the bolometric accretion luminosity from the mini-disks \citep{MM2008,DOrazioetal2013}.
However, this is unlikely to occur because most of the luminosity from the mini-disks is made in their inner radii, and the inflow time from the rim of a mini-disk is much longer than the binary orbital period \citep{Farrisetal2014}.   On the other hand, \citet{Roedigetal2014} pointed out that the shock created where streams from the circumbinary disk hit the outer rim of the mini-disks should be quite bright, radiating primarily in the hard X-ray band, and, unlike the accretion luminosity, its output should reflect the modulation because the local Compton cooling time is short.    \citet{Nobleetal2012} identified another possible signature in thermal emission from the inner edge of the circumbinary disk due to shocks driven by returning streams. The radially integrated luminosity of this component is consistent with the amount of work done by the binary torque on the disk, as argued in \citet{Shietal2012}.  Recently, however, \citet{Farrisetal2014} have computed thermal emission from the gap proper as well as its edges based on their 2D hydrodynamic simulations, reaching the surprising conclusion that, far from being dim, the streams can actually be more luminous than the inner region of the circumbinary disk.

Unfortunately, this last effort depended upon an assumption with potentially significant implications for the result.  Their heating rate was calculated on the basis of a phenomenological ``$\alpha$ model" shear viscosity.   This model was invented by \citet{ShakuraSunyaev73} to mock up internal disk accretion stresses now known to be due to MHD turbulence driven by the magneto-rotational instability \citep{BalbusHawley98}.   Because the dynamical state of the streams is quite different from the nearly-circular orbits inside an accretion disk, it is not at all clear whether this model is applicable to them.  

In this paper we set out to test these assumptions by measuring the heating rate in the streams as determined by a pair of 3-d MHD simulations first reported in \citet{ShiKrolik2015}.   These simulations make no phenomenological assumptions relevant to dissipative processes, and should therefore provide an accurate measure of local heating whether due to small-scale turbulence or coherent dissipative features like shocks.   Despite their assumed isothermal equation of state, we can also evaluate $pdV$ temperature changes from their stored velocity data.

\section{Methods}\label{sec:methods}
Our data are from the simulations of \citet{ShiKrolik2015}. Two 3-d MHD simulations of circumbinary disks around circular orbiting binaries of mass ratio $q=1$ and $0.1$ are analyzed in this work. The simulations adopted a global isothermal equation of state.  Shocks are captured with von Neumann-Richtmyer explicit bulk viscosity defined as \citep{StoneNorman1992a}
\begin{align}
  \nush \equiv
  \begin{cases} 
     \,\,\Csh {\Delta x}^2 (-\nabla\cdot \mathbf{v})  & \text{if } \nabla\cdot \mathbf{v} < 0 \\
     \,\, 0                                            & \text{otherwise, }
  \end{cases}
\label{eq:bulk_visc}
\end{align}
where $\Delta x$ is the cell width, and $\Csh = 2$ is a fixed number characterizing the number of cells over which the artificial bulk viscosity spreads a shock. 
The corresponding dissipation rate can be integrated over the volume of the gap region via:
\begin{eqnarray}
\Qsh &\equiv &\int \! rdrd\phi dz\, [\rho\nush(\nabla\cdot{\mathbf{v}})^2] \\\nonumber
     &  =    & \int\! rdrd\phi \,\Sigma\,\langle\nush(\nabla\cdot{\mathbf{v}})^2\rangle_{\rho,z} \\ \nonumber
     &  =    & \int\! rdrd\phi \, \qsh .
\label{eq:Qshock}
\end{eqnarray}
Here we define the rate of shock dissipation per unit surface area as
\beq
\qsh \equiv \Sigma\,\langle\nush(\nabla\cdot{\mathbf{v}})^2\rangle_{\rho,z} \,.
\label{eq:qshock}
\enq
This vertically-integrated quantity is useful to study because the principal dynamics of the streams are in the equatorial plane.

\begin{figure*}[!ht]
 \includegraphics[width=\columnwidth]{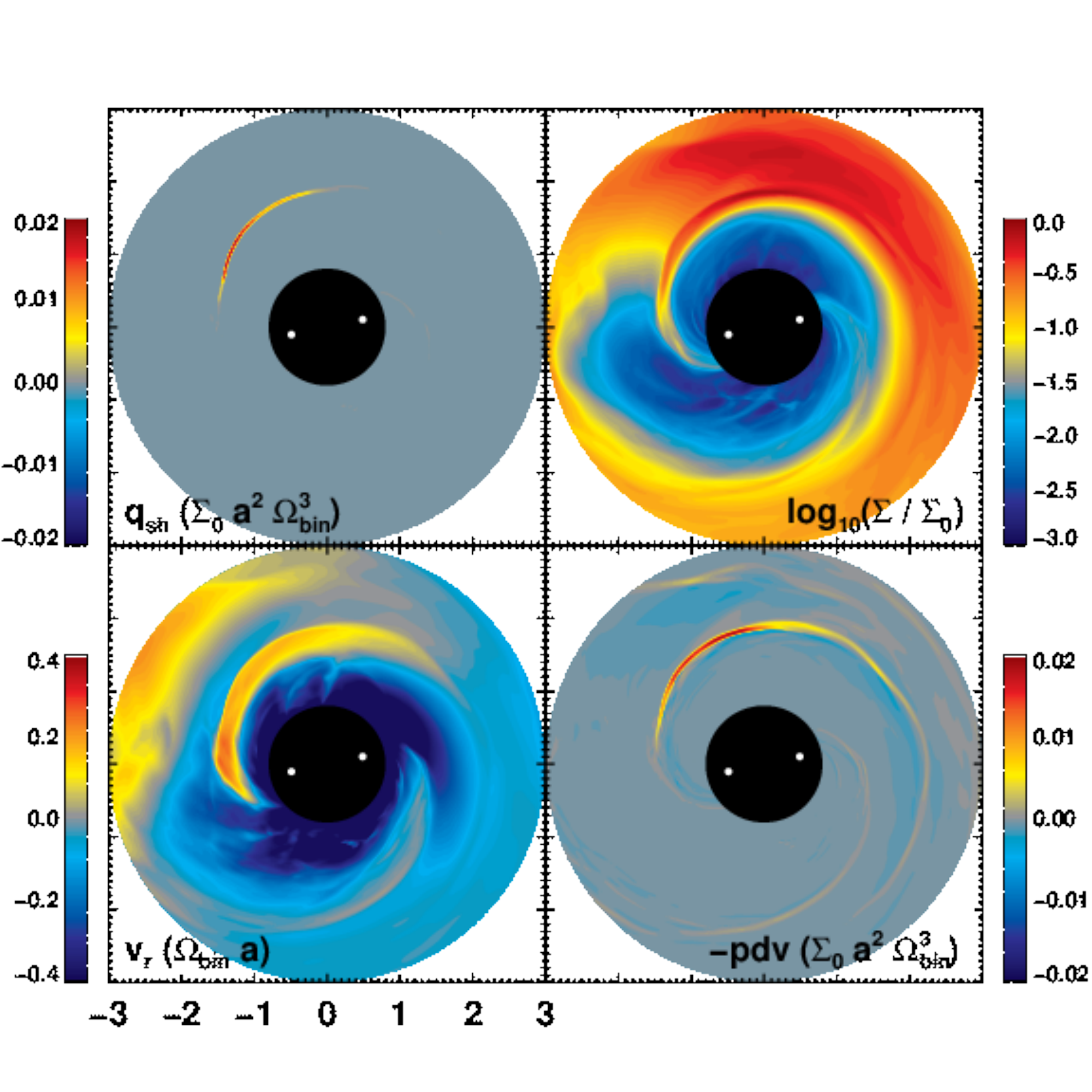} \hfill
 \includegraphics[width=\columnwidth]{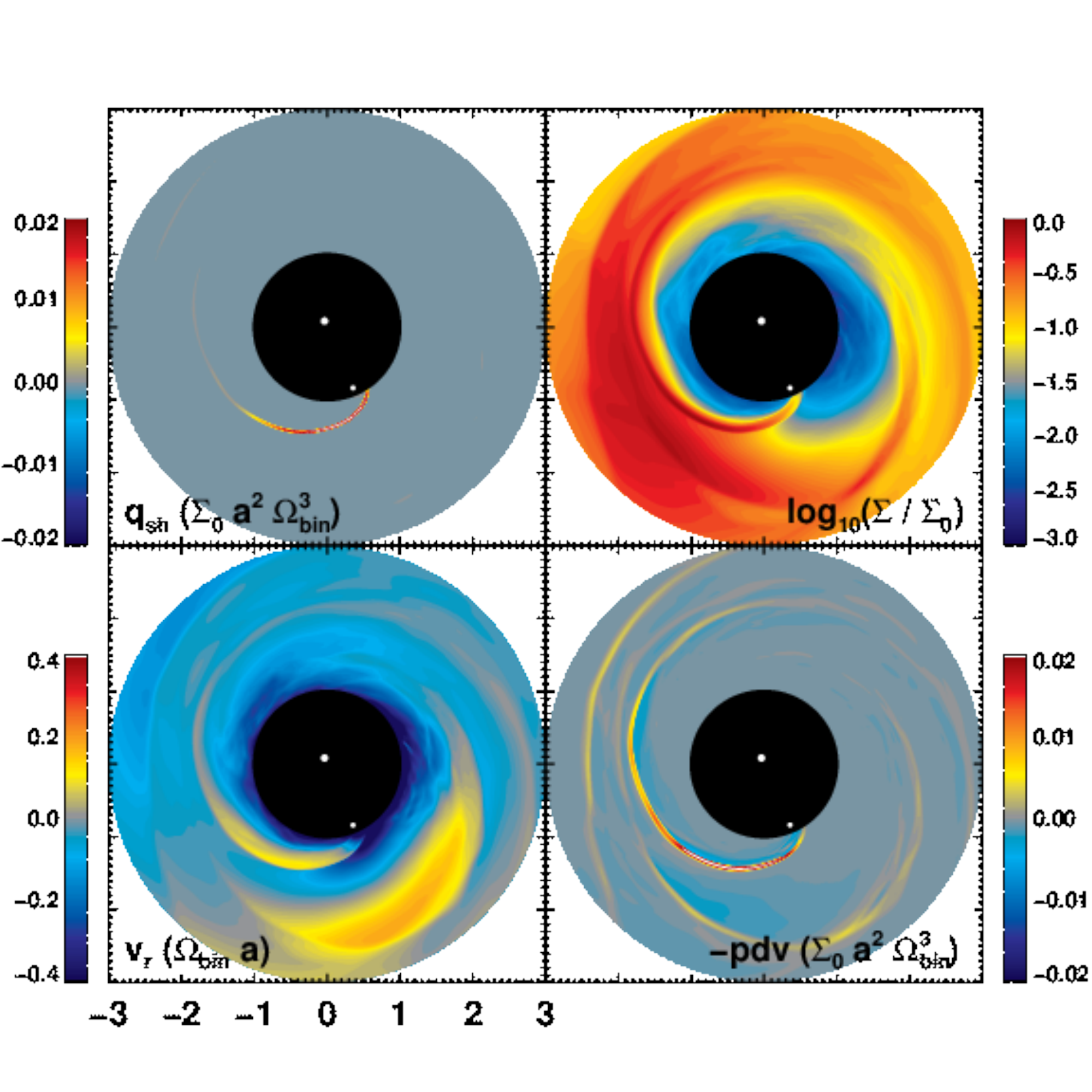}
 \caption{Snapshots of vertically-integrated shock dissipation rate per unit area ($\qsh$), surface density ($\Sigma$), density-weighted, vertically-averaged radial velocity ($v_r$), and vertically-integrated thermal work ($-pdV$) for equal mass (left four panels at $t=1329\omgbi$) and $q=0.1$ case (right four panels, at $t=1356\omgbi$).}
 \label{fig:snapshot}
\end{figure*} 
%

%pdv work
Our isothermal equation of state implicitly assumes that work done by adiabatic compression is immediately radiated away.  However, it is also equally true that an isothermal equation of state {\it injects} heat during adiabatic expansion.   The net of these two effects (if positive) is the fairest estimate of the actual luminosity associated with thermal work.  We can calculate its magnitude by evaluating the $-p\nabla\cdot{\mathbf{v}}$ term in the hydrodynamic energy equation.

\section{Results}\label{sec:results}

\subsection{Dissipation rate}\label{sec:dissipation_rate}
% A. heating concentrated in the returning stream
In Figure~\ref{fig:snapshot}, we show snapshots of the dissipation rate per unit surface area in the horizontal disk plane. The shock heating ($\qsh$) is mostly confined within the gas stream and, as we will show later, is strongest when binary torques acting on the stream push it outward.

\begin{figure*}[!ht]
 \includegraphics[width=\columnwidth]{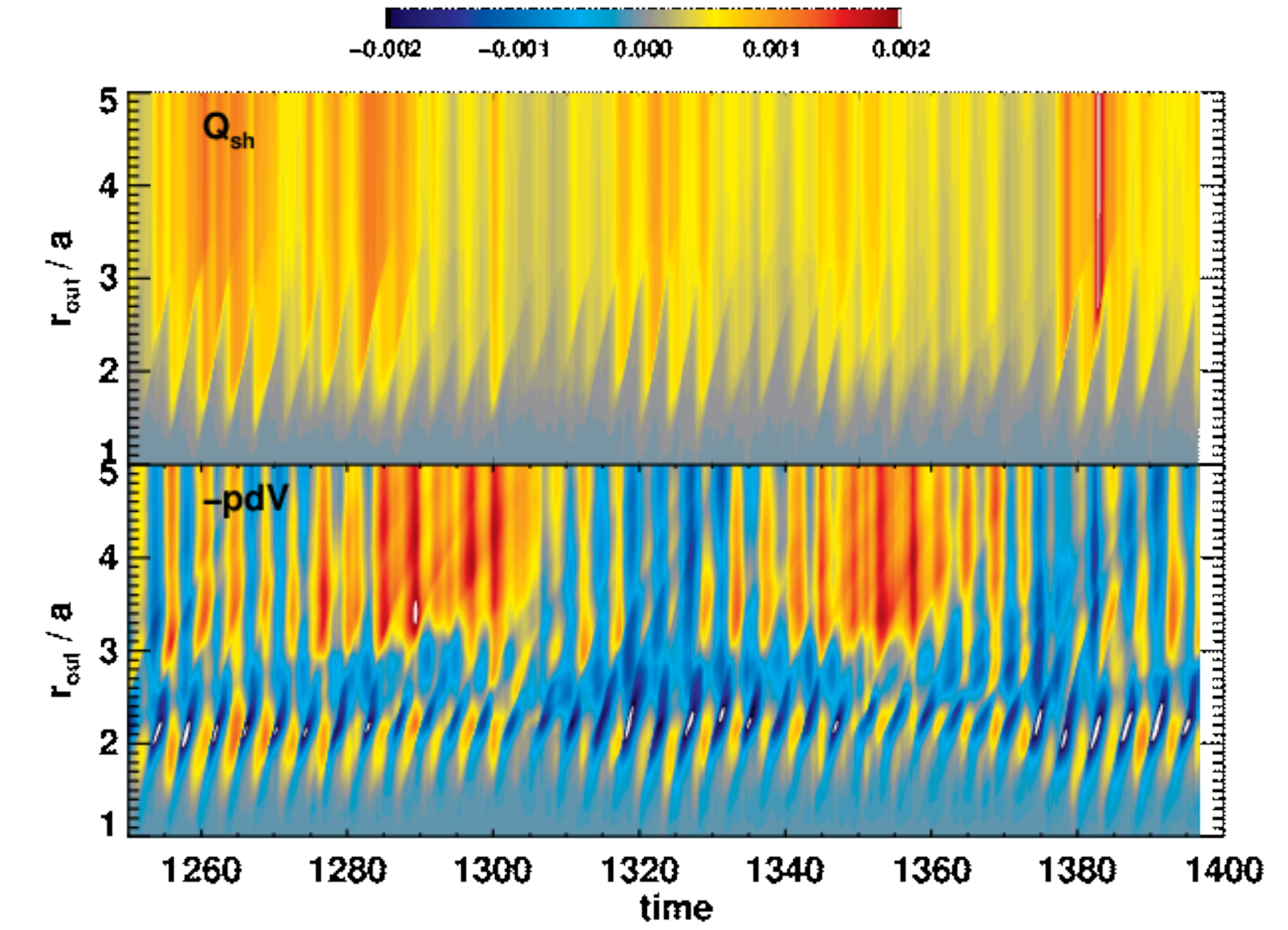} \hfill
 \includegraphics[width=\columnwidth]{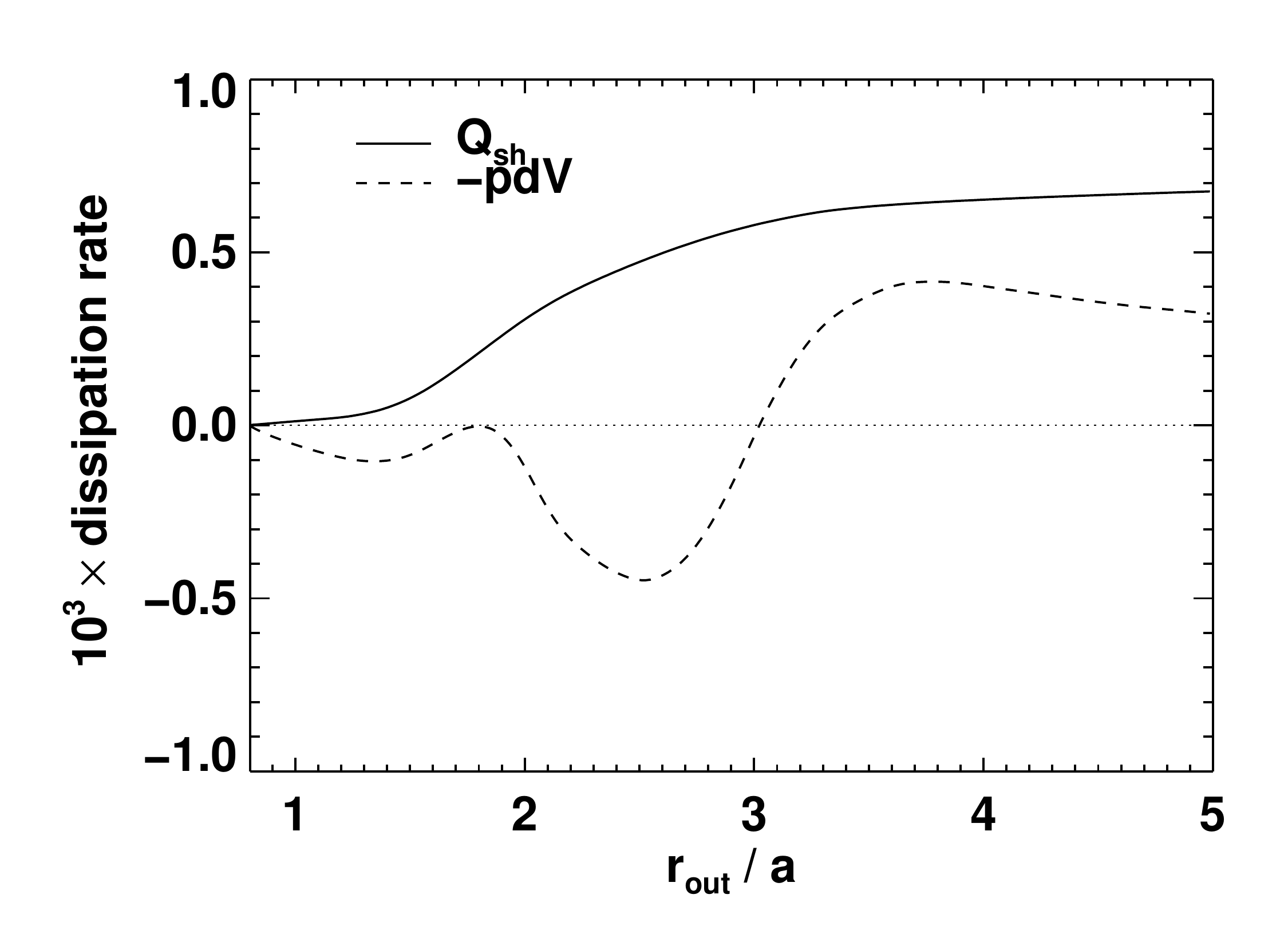} \hfill
 \includegraphics[width=\columnwidth]{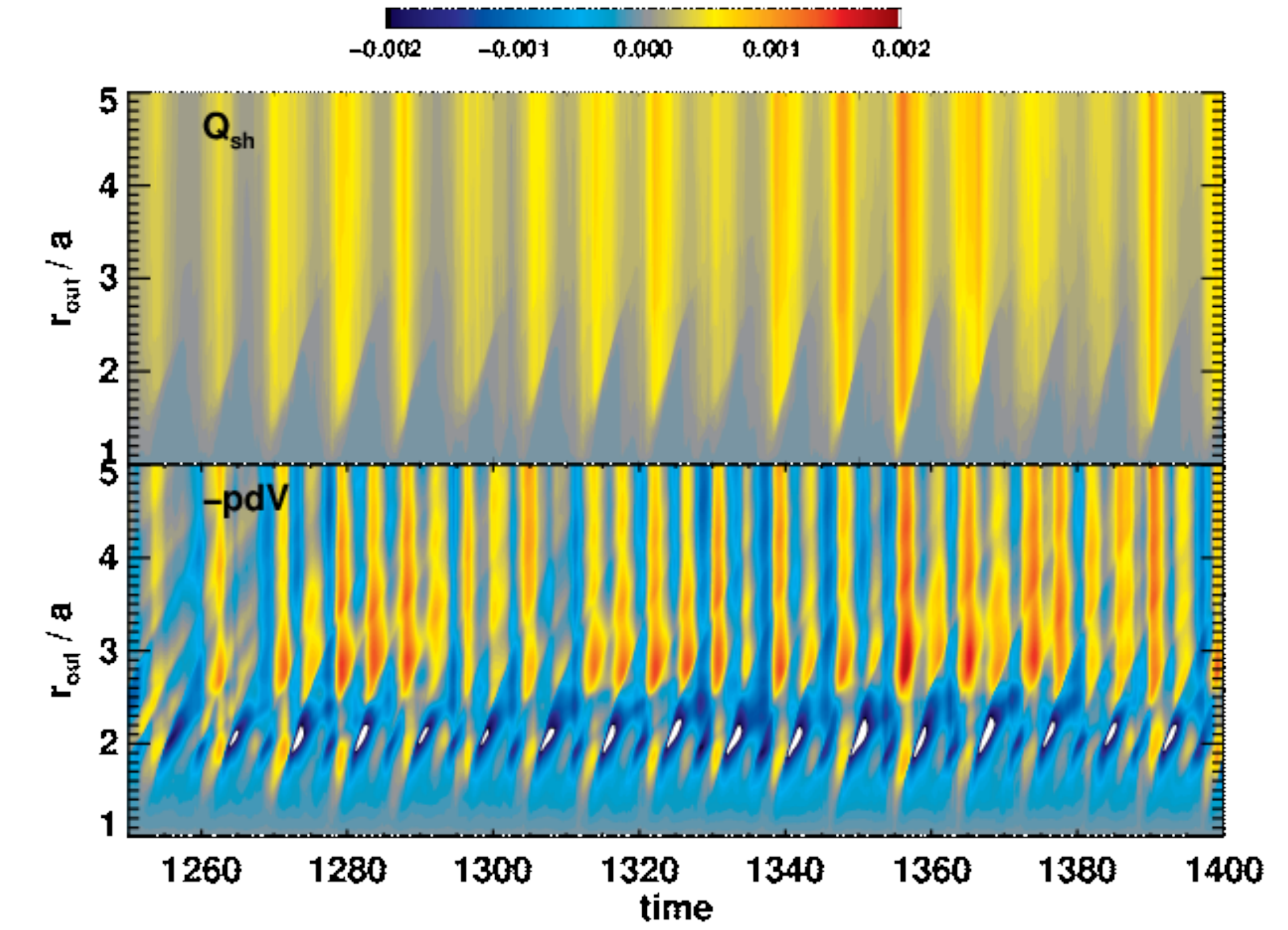} \hfill
 \includegraphics[width=\columnwidth]{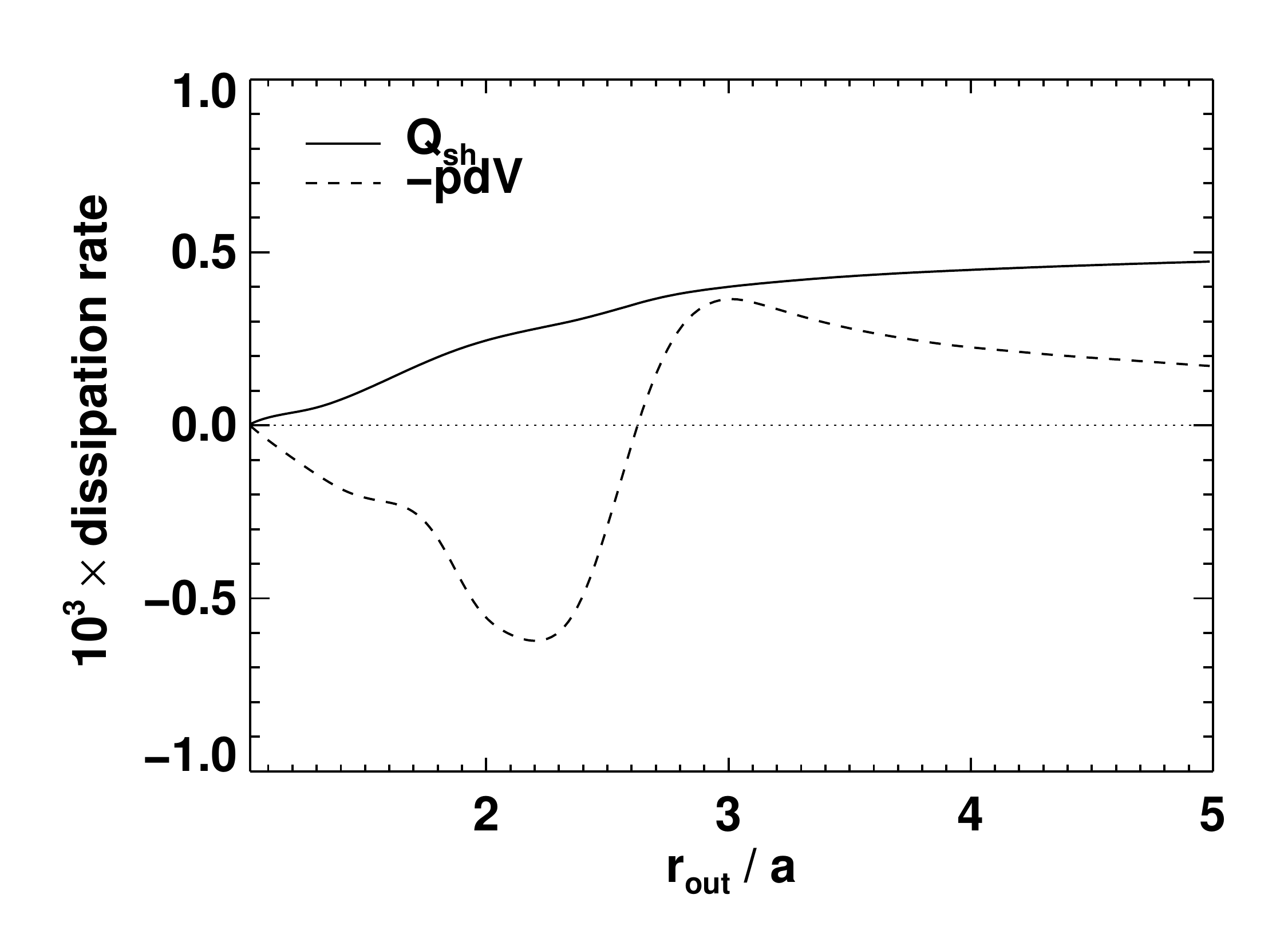}
 \caption{Integrated dissipation rates within radius $r_{\rm out}$ (in code units) as a function
 of $r_{\rm out}$ and time are shown in terms of space-time diagrams.
 Time-averaged values are shown on the right. Top two panels are for 
 $q=1$, bottom two for $q=0.1$.}
 \label{fig:diag}
\end{figure*}

% choice of the rout:
In Figure~\ref{fig:diag}, we display the dissipation rate integrated out to $r_{\rm out}$ as a function of time.
As can be seen, shocks are well-defined events, so we can easily choose an outer boundary for their contribution to 
the heating. By definition, the shock heating $\Qsh$ is always positive. 
In the equal mass binary, the time averaged shock heating takes places mostly around $r\gtrsim 1.5 a$, where the shocks are strongest. A smaller amount takes place at $r\sim 3 a$ where the shocks propagate into the disk interior.  On the other hand, in the $q=0.1$ binary, the time averaged heating rate is distributed more broadly in radius over the range $a \lesssim r \lesssim 3 a$.

In contrast, the $-pdV$ work is somewhat more broadly distributed spatially, and can have either sign.  Although its absolute magnitude can be greater than that of $q_{\rm sh}$, its variation in sign leads to substantial cancellation.
In fact, Figure~\ref{fig:diag} shows that the net time-averaged $-pdV$ work is always negative, i.e., creates {\it cooling}, within the gap region. Indeed, it is particularly large and negative at the inner edge of the circumbinary disk, just inside the surface density maximum \citep[see also Figure~4 of ][]{ShiKrolik2015}.  For our volume-integrated quantities, we therefore choose the gap's outer edge $r_{\rm out} = 2.5 a$ for the equal mass binary and $2.2 a$ for the $q=0.1$ case.
%{\color{red} Fortunately, o
We note that our qualitative conclusions, both about the net effect of adiabatic temperature changes and the overall magnitude of stream luminosity, do not depend in any significant way on the exact choice of $r_{\rm out}$.

%C. time-averaged luminosity $\sim$ few percent $\times GM\dot{M}/a$, $ll$ \citep{Farrisetal2015}

\begin{figure}[!ht]
 \epsscale{1.0}
 \plotone{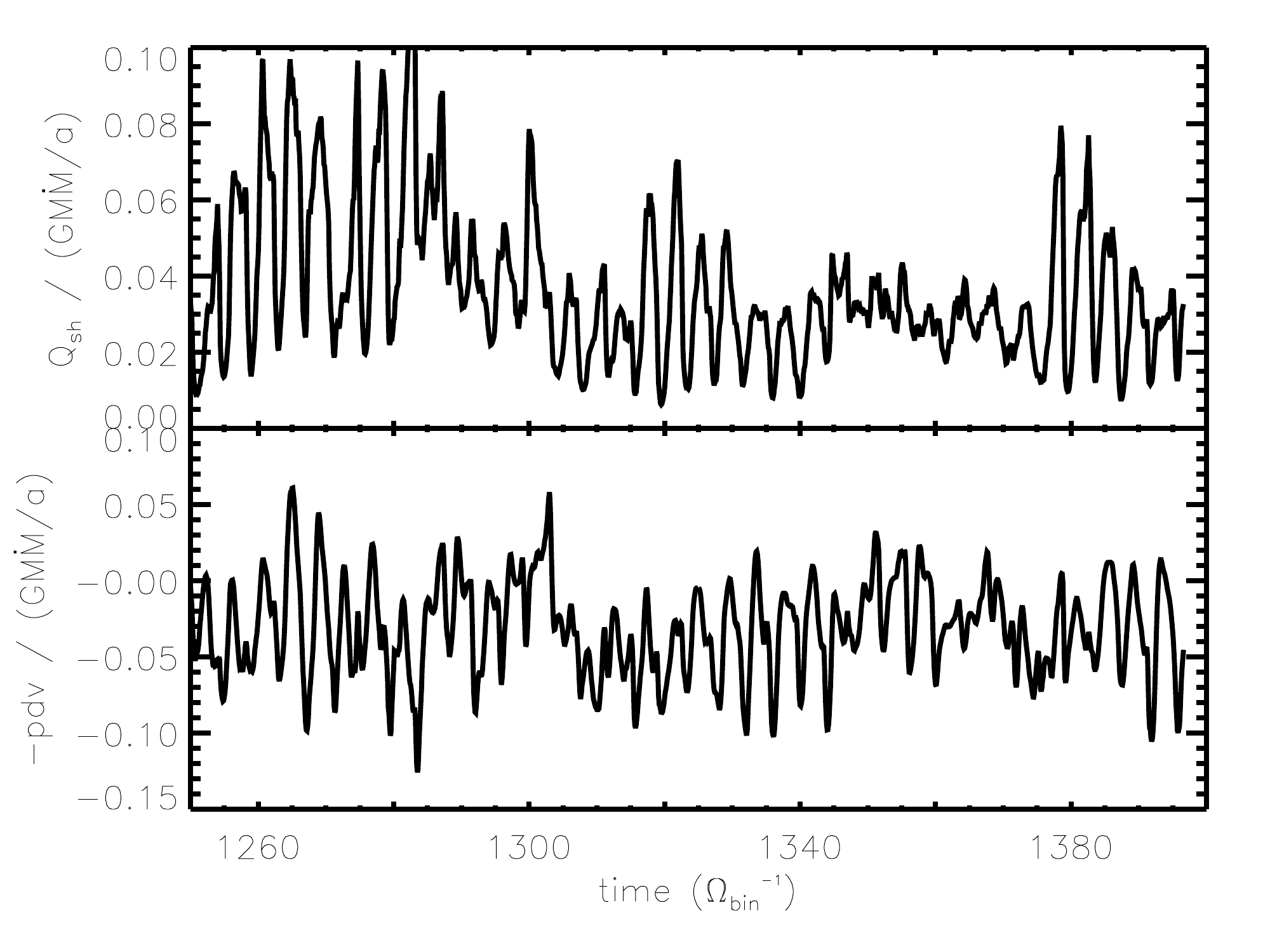} \\
 \plotone{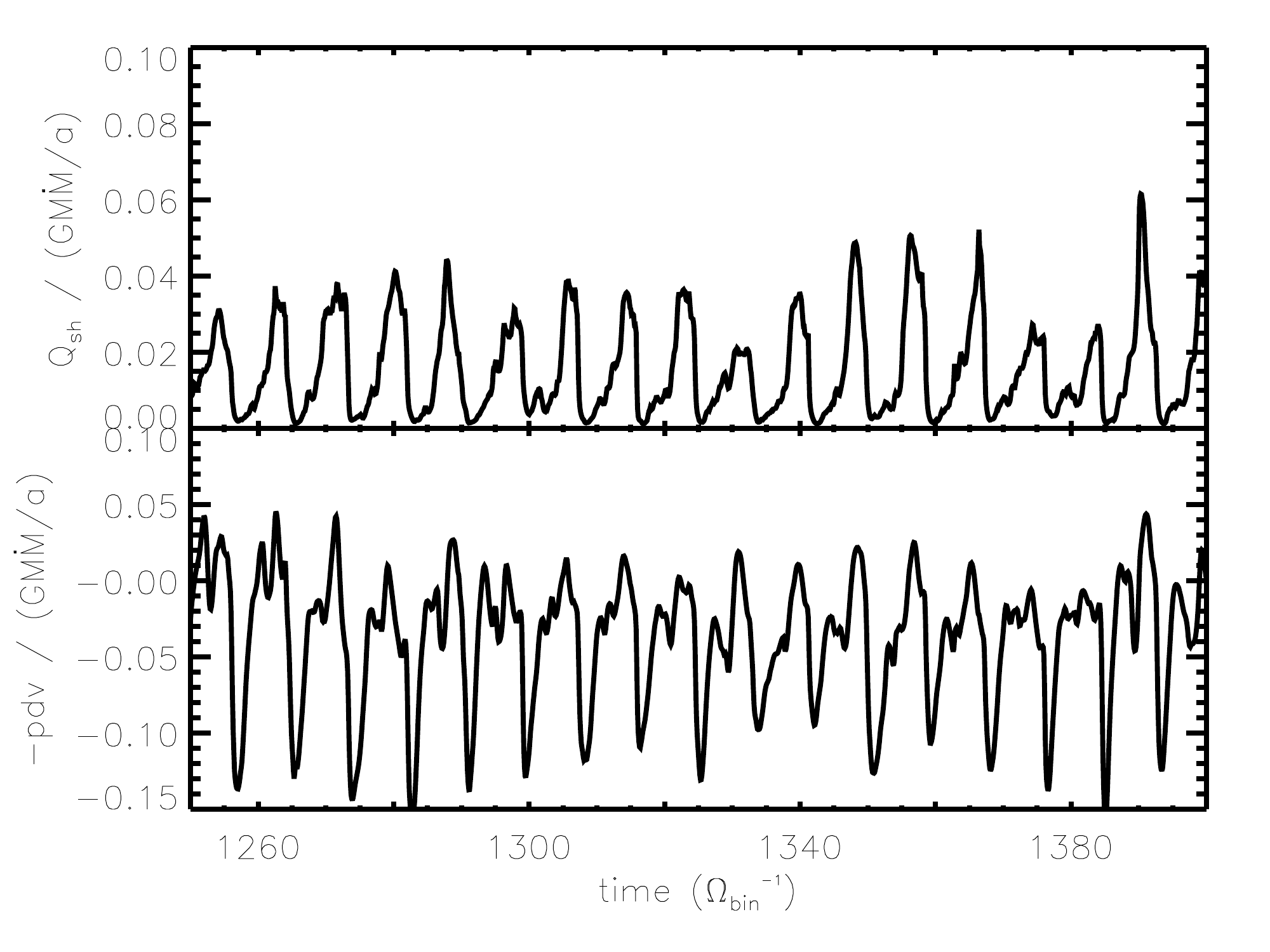}
 \caption{Time history of $\Qsh$ and $-pdV$ normalized to the instantaneous  $GM\dot{M}/a$ (at the inner boundary) for $q=1$ (top two  panels) and $q=0.1$ (bottom two panels).}
 \label{fig:history}
\end{figure}

We plot the time history of the radially-integrated dissipation rates in Figure~\ref{fig:history}, normalizing the dissipation to the instantaneous value of $GM\dot{M}/a$. The accretion rate is measured at the inner boundary $r_{\rm in}$, and we neglect the small difference between $\dot{M} (r=r_{\rm in})$ and $\dot{M} (r=a)$.  We find: (1) 
$\Qsh$ and $-pdV$ are comparable to one another, but in this volume-integrated sense $-pdV$ is a factor of a few smaller and generally negative. (2) both $q=0.1$ and $q=1$ show strong oscillations even after the normalization; the $q=0.1$ case shows more regular and lower frequency fluctuations (when measured in units of $\omgbi$) although the amplitude is in general smaller than in the equal mass case. (3) in both cases and at all times, both heating rates are less than $10\%$ of the total power that could be tapped from accretion; in an average sense, their magnitudes are closer to a few percent of that power. To be quantitative, the actual time-averaged normalized heating rates are: $\langle\Qsh/(GM\dot{M}/a)\rangle_{\rm t} \simeq 0.037$ ($0.017$) for $q=1$ ($q=0.1$), $\langle -pdV/(GM\dot{M}/a) \rangle_{\rm t} \simeq -0.031$ ($-0.038$) for $q=1$ ($q=0.1$).
%%%%%%%%%%%%%%%%%%%%%%%%%%%%%%%%%%%%%%%%%%%%%%%%%%%%%%%%%%%%%%%%%%%
% q = 1  rout=2.5 // time avg over t=[    1250,      1396.60]:
%qdiss =   0.000474184 mdot =     0.0137242 
%qdiss/(GMMdot/a) =     0.0365648 -PdV/(GMMdot/a) =    -0.0308505
% q =0.1 rout=2.2 // time avg over t=[    1250,      1400.00]:
%qdiss =   0.000278708 mdot =     0.0167368 
%qdiss/(GMMdot/a) =     0.0169315 -PdV/(GMMdot/a) =    -0.0380668
%%%%%%%%%%%%%%%%%%%%%%%%%%%%%%%%%%%%%%%%%%%%%%%%%%%%%%%%%%%%%%%%%%%%

\begin{figure}[!ht]
 \includegraphics[width=\columnwidth]{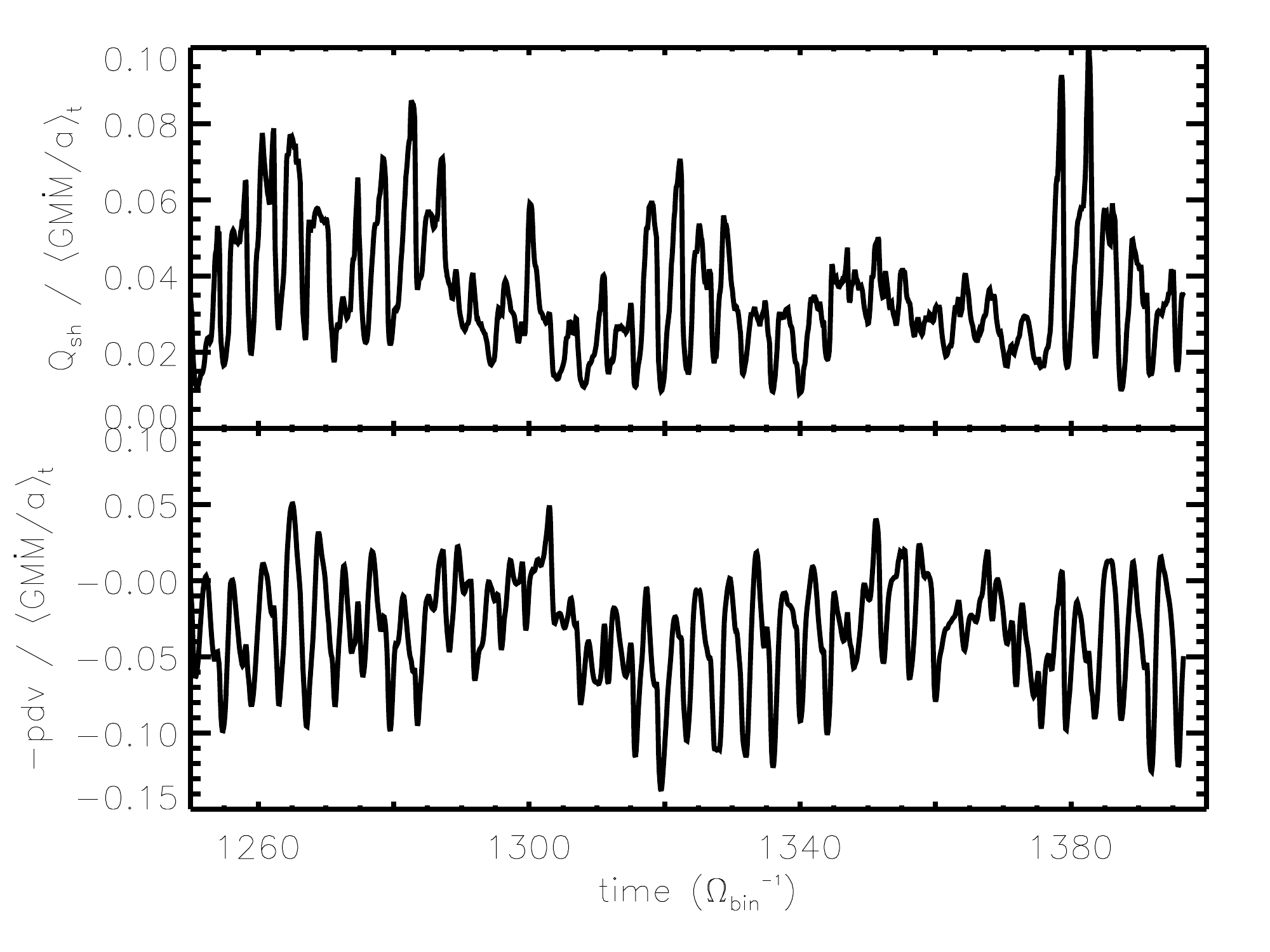} \\
 \includegraphics[width=\columnwidth]{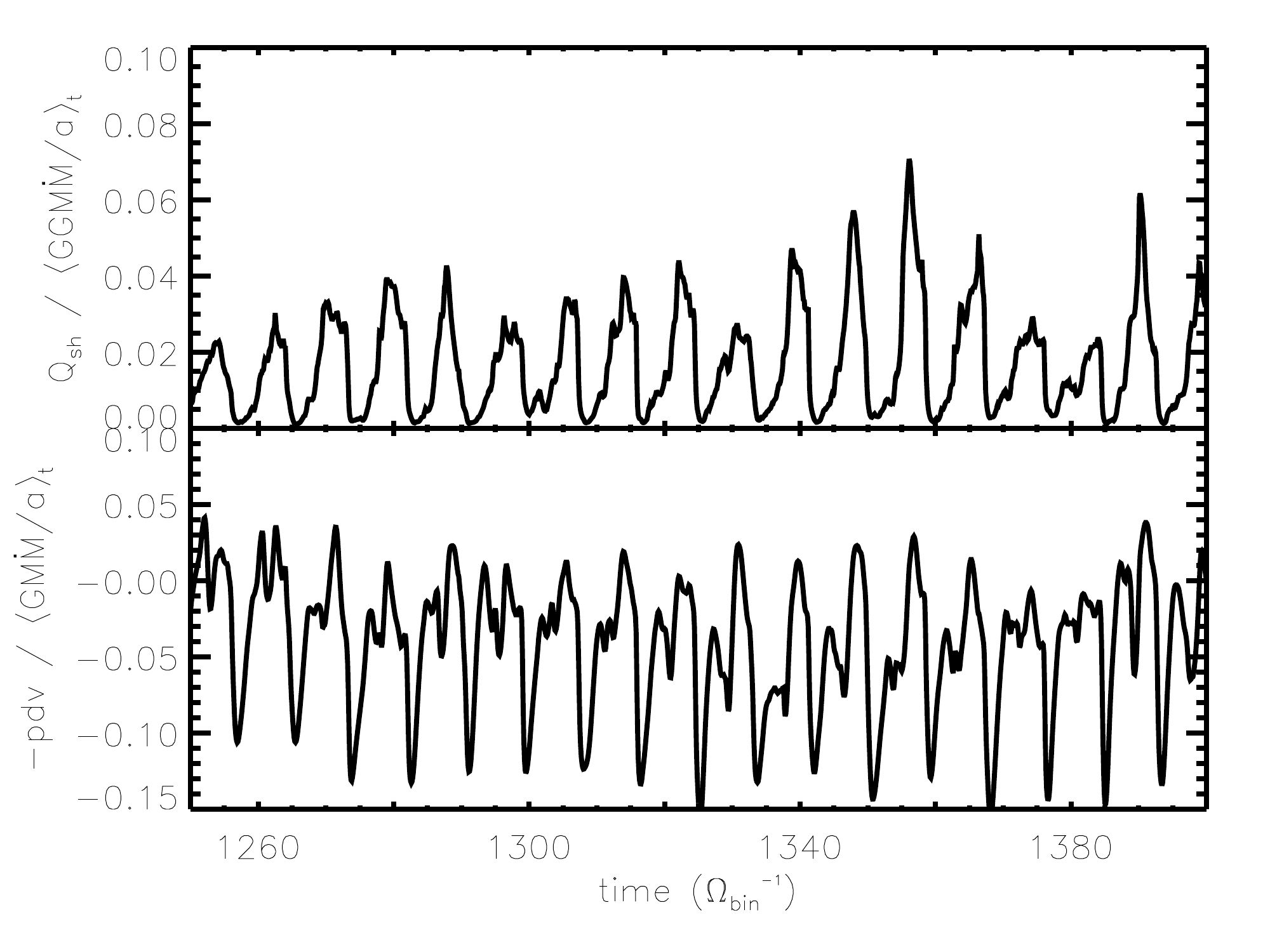}
 \caption{Same as Figure~\ref{fig:history} but normalized to the time averaged $\langle GM\dot{M}/a\rangle_{\rm t}$ measured at the inner boundary.}
 \label{fig:history_nonorm}
\end{figure}

Figure~\ref{fig:history_nonorm} shows the same evolution of dissipation
but normalized with time averaged accretion 
$\langle GM\dot{M}/a\rangle_{\rm t}$. Surprisingly, the amplitudes
change rather little.  The reason is that the heating rates are
modulated at the same frequencies as the accretion rate, but with a
phase offset.   This offset provides an important clue to the mechanism
driving the shocks.

\begin{figure} [!ht]
 \includegraphics[width=\columnwidth]{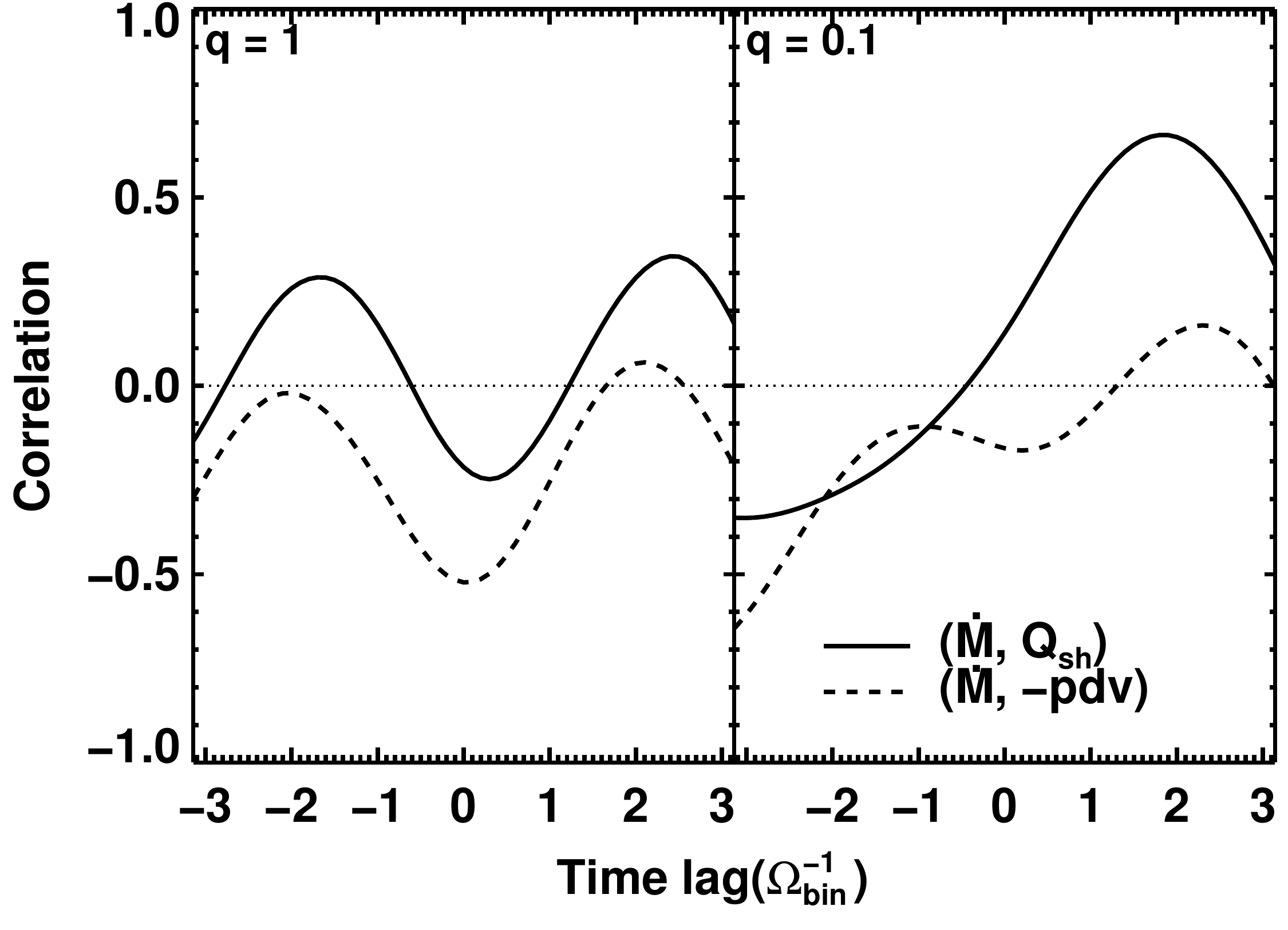}
 \caption{Cross correlation between the accretion rate and the dissipation rates. Shock heating variations follow accretion rate variations with a lag $\sim 2\omgbi$\,.(black solid lines)}
 \label{fig:lag}
\end{figure}

To study this phase relationship, we compute the cross-correlation functions between the heating rates and the accretion rate.  These are shown in Figure~\ref{fig:lag}. When $q=0.1$, there is a strong peak in the cross-correlation function indicating that fluctuations in the shock heating rate follow fluctuations in the accretion rate with a lag $\simeq 2\omgbi$.  That mass-ratio also displays a weaker, but significant, peak at a slightly larger lag for fluctuations in the adiabatic compression work.  Two peaks appear in the $q=1$ case, but they are actually the same peak: both the accretion rate and the shock heating rate are modulated at a frequency $\simeq 1.5\omgb$, whose period ($\sim 4\omgbi$) is the separation between the two peaks.   Thus, in this case, too, the heating rates follow variations in the accretion rate with a lag $\simeq 2\omgbi$. This delay occurs because the shocks take place in matter that has fallen in almost to the inner cut-out, but then moves outward as binary torques raise its angular momentum.  The outward acceleration takes $\simeq 2\omgbi$ to be accomplished.

\subsection{Gridscale dissipation}\label{sec:grid_diss}
\begin{figure}[!ht]
 \includegraphics[width=\columnwidth]{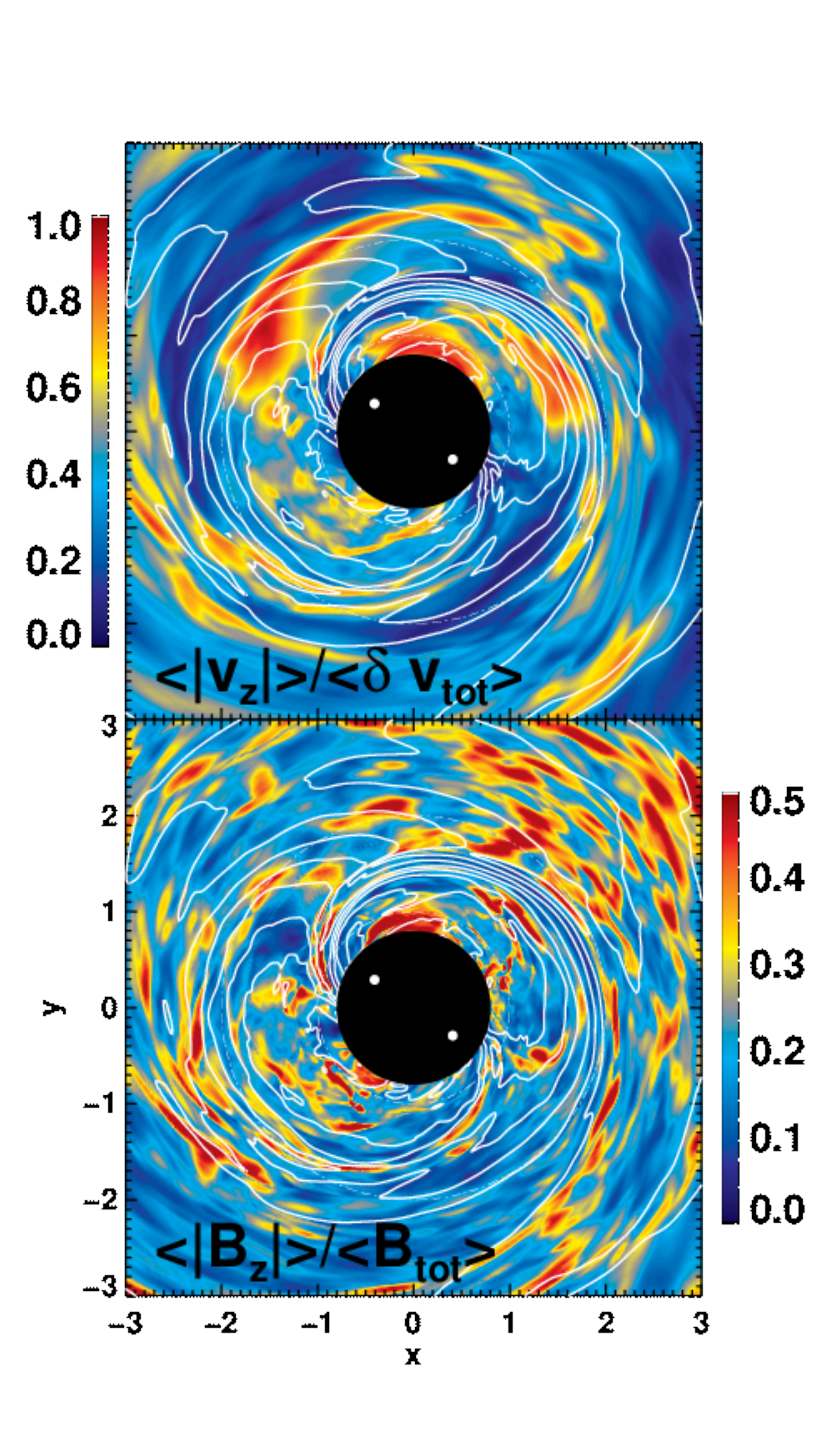}
 %\plotone{a_turb.png}
 \caption{
 $\langle |\vz|\rangle_{\rho,z}/\langle (v_{\rm r}^2+\delta v_{\phi}^2+\vz^2)^{1/2}\rangle_{\rho,z}$ (top) and $\langle |\Bz|\rangle_{\rho,z}/\langle \mathbf{B}^2 \rangle_{\rho,z}$ (bottom) 
 in color scales which characterize the amplitude of the turbulence. The overlaid surface density contours (in white) show ten consecutive
 values logarithmically scaled between $10^{-3}$ to $1$.  Clearly, 
 within the gas stream (where contour lines concentrate) the turbulent amplitude is largely weaker than that in the disk body.}
\label{fig:aturb}
\end{figure}
So far, we have identified sources of luminosity as either genuine shock heating or adiabatic compression.  However, in turbulent fluids there can also be a contribution from subsonic dissipation acting on the scale of the smallest turbulent eddies, creating heat by dissipation of either fluid kinetic energy or magnetic fields.  This kind of dissipation provides most of the heating in ordinary accretion disks;  it is also the sort of heating that the $\alpha$-model hopes to capture.

It is, however, very difficult to capture in a realistic way in fluid simulations because the physical dissipation lengthscale is very very small compared to feasible grid resolutions.   Instead, numerical simulations mimic this dissipation by gridscale numerical effects. Because MHD simulations can develop turbulence as a result of the magnetorotational instability, there is a reasonable hope that the turbulent cascade in such a simulation carries energy to the dissipation scale at the same rate as in a physical disk.  Hydrodynamic disks do not exhibit turbulence; hence the insertion of phenomenological models to create dissipation.  In energy-conserving (and non-isothermal) MHD simulations, the gridscale dissipation is automatically captured into heat. However, the $\texttt{ZEUS}$ code used in these simulations solves an internal energy equation rather than a total energy equation, and our equation of state is isothermal.  Nonetheless, it is still possible to estimate the magnitude of this sort of dissipation in our simulations by supposing that the shear part (the traceless part) of the velocity gradient tensor is comparable in magnitude to the divergence part (the part containing non-zero trace), and then supposing that the numerical effective shear viscosity is comparable to the explicit artificial bulk viscosity. 

In the bulk of the circumbinary disk, turbulent dissipation almost certainly dominates the local heating rate.   However, the gap region is different. The streams are nearly laminar, i.e., their fluctuations in fluid quantities are small compared to their mean values. We quantify this statement by defining the fractional turbulent velocity amplitude as $\langle |\vz|\rangle_{\rho,z}/\langle (v_{\rm r}^2+\delta v_{\phi}^2+\vz^2)^{1/2}\rangle_{\rho,z}$ and the fractional turbulent magnetic field amplitude as $\langle |\Bz|\rangle_{\rho,z}/\langle \mathbf{B}^2 \rangle_{\rho,z}$.  The values of these quantities are shown in Figure~\ref{fig:aturb}.  The accretion streams' locations within these images are identified by the concentrated white contours, which show the surface density.   Within the streams, the fractional turbulent amplitudes of both velocity and magnetic field are clearly much smaller than in the truly turbulent circumbinary disk: whereas both amplitudes are frequently $\simeq 0.5$ in the circumbinary disk, they are both almost always $< 0.1$ in the accretion streams, and frequently closer to 0.01.   Because dissipation scales with the square of the fluctuation amplitude, we estimate that turbulent dissipation within the streams must be quite small compared to the shock heating, almost certainly $< 10^{-2} Q_{\rm shock}$.  In the streams, therefore, turbulent dissipation adds little heating to the already small amount created by shocks.

\subsection{Thermal spectra} \label{sec:sed}
We now calculate the thermal spectrum of an accreting binary by taking account shock heating in the streams crossing the gap as well as accretion heating in both the circumbinary disk and the mini-disks.  For the two conventional disk regions, we follow
\citet{Roedigetal2014}. 
We define a characteristic temperature by assuming steady accretion and radiative cooling
locally at $r=a$:
\beq
T_0 \equiv \left(\frac{3GM\dot{M}(a)}{8\pi\sigma a^3}\right)^{1/4}\,,
\label{eq:T0}
\enq
where $\dot{M}(a)$ is the time averaged accretion rate at $r=a$. \footnote{For the $q=0.1$ 
case, we neglect the small difference between $r=a$ and the inner simulational boundary,
which is at $r\simeq 1.02$.}
The effective temperature at any radius within either the circumbinary disk or the mini-disks would scale as 
\beq
{T^{\rm(disk)}_{\rm eff}}(r)\, /\, {T_0} = \left(\frac{\dot{M}(r)}{\dot{M}(a)}\right)^{1/4} \left(\frac{a}{r}\right)^{3/4},
\label{eq:teff_disk}
\enq
where $r$ is the distance from the system center-of-mass in the circumbinary disk, while it is the distance from the corresponding black hole in a mini-disk.  The accretion rate is likewise divided into individual mini-disk portions.   Our simulation achieved steady accretion out to to $r\lesssim 4 a$, so the local effective temperature in the circumbinary disk is mostly
determined by the power law $r^{-3/4}$. We further assume the two mini-disks are filled to their tidal truncation radii. The temperature at the outside edges of individual
mini-disks are calculated with equations~(4-5) in \citet{Roedigetal2014}, which assume the two mini-disks receive fractions $f_{1,2}$ of the total accretion rate, where $f_1+f_2=1$ and $f_1/f_2 = q$.

Following \citet{Roedigetal2014}, we write the spectral emissivity of all the disks, circumbinary or mini, in the form
\beq
L^{\rm(disk)}_{\epsilon} = \frac{32 \pi a^2}{3 h^3 c^2 g^4} (k T_0)^{8/3} 
                            \epsilon^{1/3}\!\!
                            \int^{u_h}_{u_l} du \frac{u^{5/3}}{e^{u/g}-1} \,,
\enq
where $\epsilon \equiv h\nu$, $u \equiv \epsilon/(k T_{\rm eff}^{\rm (disk)})$, $g\simeq1.7$ is the hardening factor due to strong electron scattering, and $u_l$
and $u_h$ are the lower and upper bounds of the integral, which are set by the effective
temperatures at the inner and outer edge of the disk.  The inner disks are integrated from $u_l=0$ up to $u1$ and $u2$ defined in Equation
(9-10) of \citet{Roedigetal2014}; the emission of the outer disk is integrated from $r=r_{\rm out}$ ($2.5a$ for the $q=1$ binary and $2.2a$ for the $q=0.1$ binary) out to $4a$.

Similarly, the effective temperature of the stream shock regions can be scaled as 
\beq
{T^{\rm(sh)}_{\rm eff}}\,/\,{T_0} = \left(\frac{8\pi a^3\qsh}{3GM\dot{M}(a)}\right)^{1/4}\,.
\label{eq:teff_shock}
\enq
As Figure~\ref{fig:snapshot} illustrates, $T^{\rm(sh)}_{\rm eff}$ is a strong function of position within the gap region.
We can then integrate over the shock heating regions and find 
\beq
L^{\rm(sh)}_{\epsilon} = \frac{8 \pi a^2}{h^3 c^2g^4}\,
                         \epsilon^3\!\!
                         \int^{r_{\rm out}}_{r_{\rm in}} \!\!\frac{r/a}{e^{u/g} - 1} 
                         d\left(\frac{r}{a}\right)\frac{d\phi}{2 \pi} \,,
\enq
where $u = \epsilon/(k T_{\rm eff}^{(\rm sh)})$. Here $r_{\rm in} = a$ and $r_{\rm out} = 2.5a$ ($2.2a$) for the $q=1$ ($q=0.1$) case, as discussed in section~\ref{sec:dissipation_rate}.

\begin{figure*}[!ht]
 \epsscale{2.5}
 \plottwo{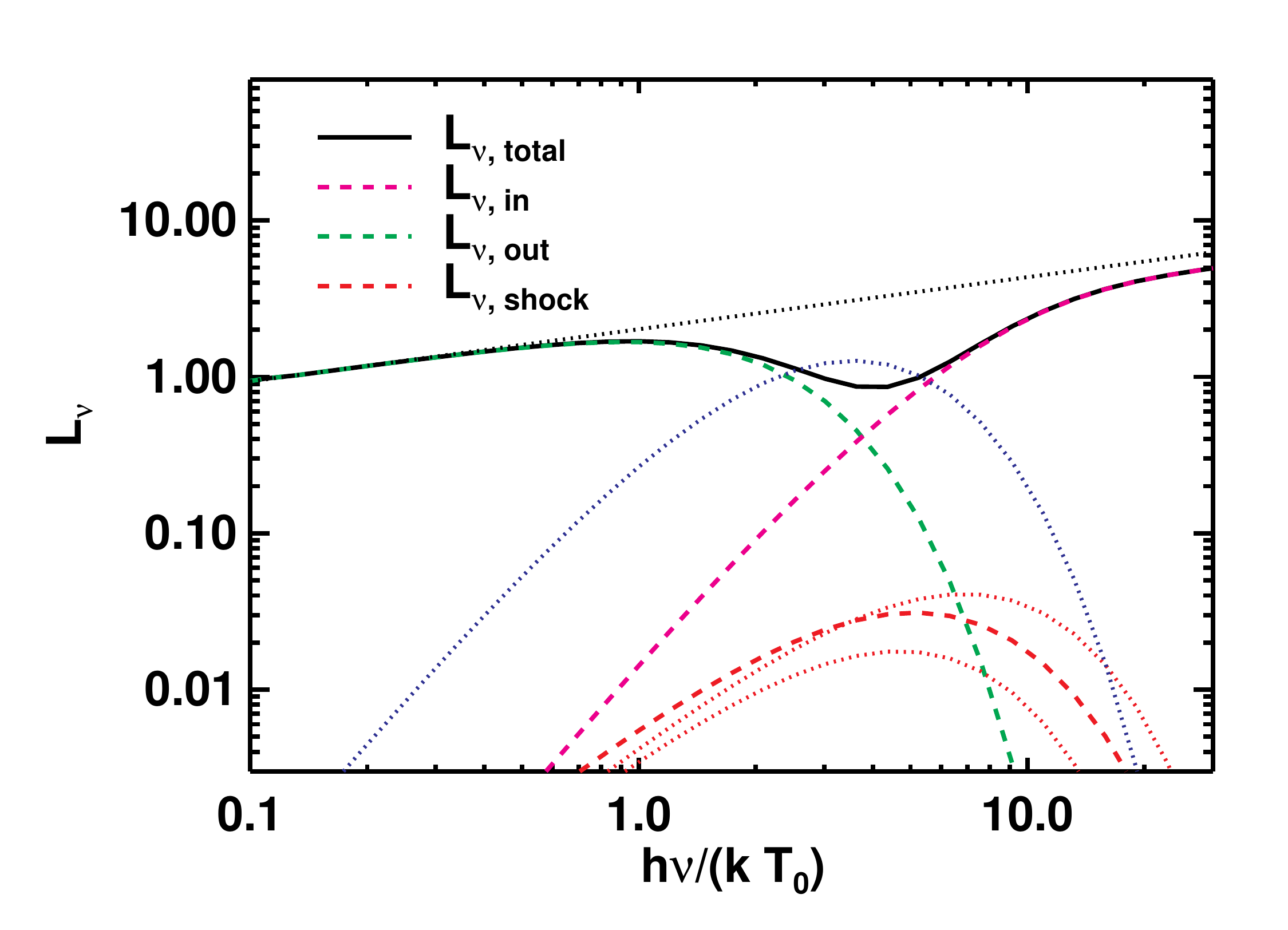}{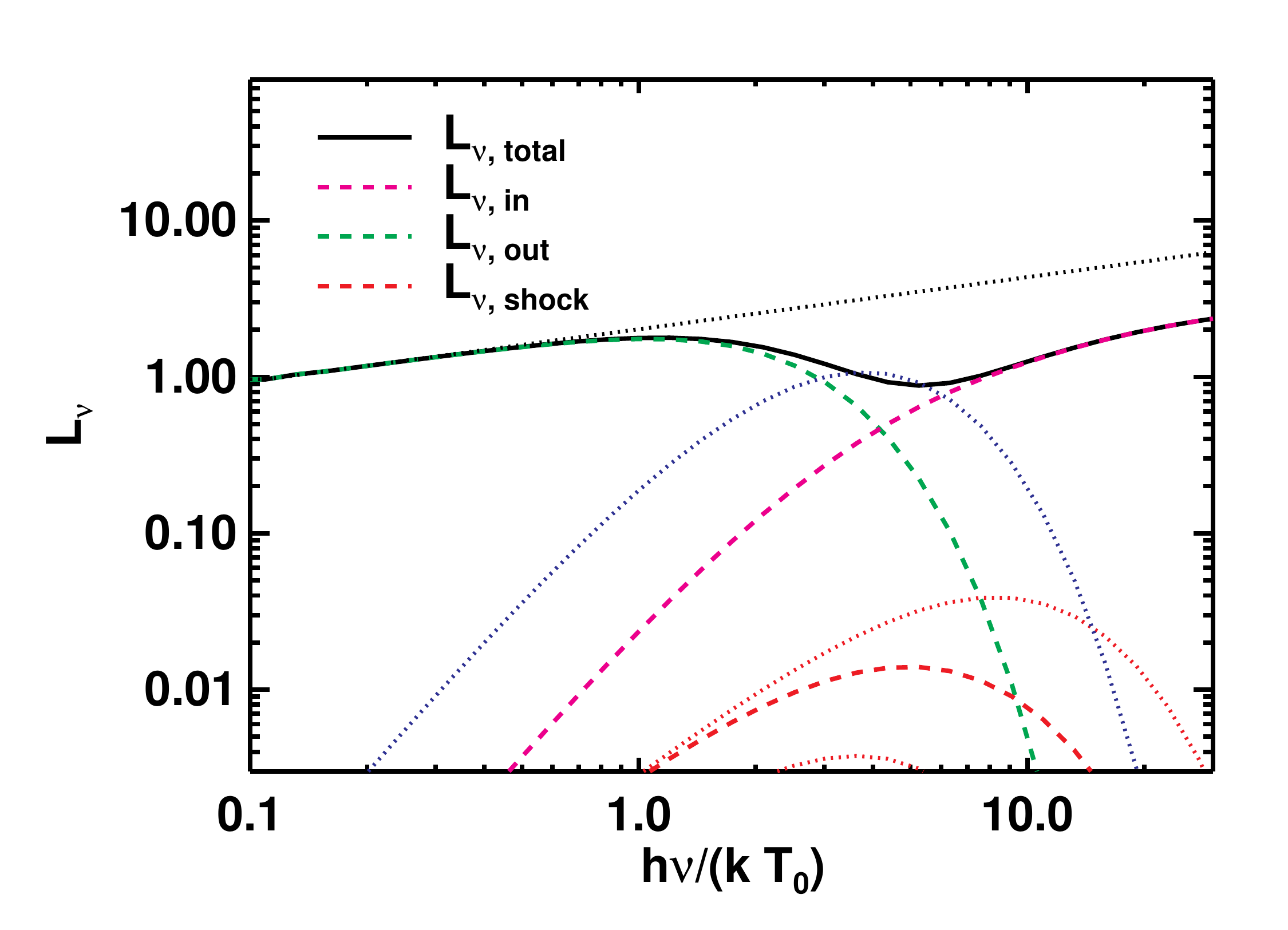}
 \caption{Thermal spectra for circumbinary disks with equal (left, $q=1$) and unequal (right, $q=0.1$) mass binaries.
 The solid black lines show the total spectra, while different components are in color as labeled in the legend.
 %For both cases, the spectral contribution from shock heating of the streams (red) is not big enough to fill the dark  ``notch" due to the low emissivity of the gap regions.
 The red dashed curves show the time averaged values of the stream; the red dotted curves are the largest and smallest
 instantaneous contributions to the luminosity. For comparison, a classic steady accretion disk component between
 $r_{\rm in}$ and $r_{\rm out}$, the gap region, is shown as a blue dotted curve. The black dotted line is the classic
 $\epsilon^{1/3}$ spectrum.}
 \label{fig:spectra}
\end{figure*}

The resulting spectra are displayed in Figure~\ref{fig:spectra} for both mass-ratio cases.  Clearly, the deep depression of the binary spectrum (black curve) is still present even after the radiation within the gap region (red dashed) is added. Compared to the radiation from the gap region that would have been emitted from a classic disk (no gap; blue dotted), the stream luminosity is two orders
of magnitude smaller in amplitude, and what little is radiated is shifted to photons roughly a factor of two higher in energy.
The time variation of the stream contribution to the spectrum is bounded by the two red dotted lines, which show the most extreme values of the luminosity found during our simulation.  Even these extrema differ by only a factor of a few from the mean values, so at no time do they make significant contributions to filling the ``notch" shown in the spectra of \citet{Roedigetal2014}.

These results differ from those of \citet{Farrisetal2015}, who found, using the $\alpha$-model, that emission from the streams can mask the ``notches" completely. It appears that the $\alpha$-model, designed to mimic turbulent dissipation through a ``viscous" stress proportional to the local shear, overestimates the heating rate because there is substantial shear in the streams, but only weak turbulence.  The actual stream luminosity is mainly due to laminar shocks, and is much smaller than the $\alpha$-model estimate.

\section{Conclusions and Implications}\label{sec:implications}

Given that shock heating in the streams is at most a few percent of the accretion luminosity on the radial scale of the gap, and our best estimate of turbulent dissipation within the streams is that it should be $< 1\%$ of the shock heating, it seems unlikely that their luminosity can be more than a rather small fraction of the ordinary accretion luminosity produced either near the inner edge of the circumbinary disk or near the outer edges of the mini-disks around the members of the binary.   For this reason, we disagree with the claim made by \citet{Farrisetal2015}, who applied an ``$\alpha$-viscosity" model to stream fluid dynamics, that the streams should, if anything, be brighter than the adjacent portions of the disks.  Instead, our results support the prediction of \citet{Roedigetal2014} that the thermal continua of accreting binaries should be marked by a ``notch" centered on the frequencies corresponding to the temperature an ordinary disk would have on a radial scale of order the binary separation.

%put the temperature etc into dimensional analysis
For typical parameters of supermassive binary black holes, the characteristic disk temperature scales as
\begin{eqnarray}
T_0 \!\!&=& \!\!1.9 \times 10^4 \left(\frac{\eta}{0.1}\right)^{-1/4}\left(\frac{\dot{M}}{\dot{M}_{\rm E}}\right)^{1/4} \!\!\! \\\nonumber
\!\!&\times &\!\! \left(\frac{M}{10^8{\rm M_\odot}}\right)^{-1/4} \!\!\!
\left(\frac{a}{10^2\,R_{\rm g}}\right)^{-3/4} \!\! {\rm K} \,,
\label{eq:T0_unit}
\end{eqnarray}
where $\dot{M}_{\rm E} \equiv L_{\rm E} /(\eta \, c^2)$, with $\eta$ denoting the accretion energy conversion rate, $R_{\rm g}$ the 
Schwarzschild radius, $\dot{M}$ the outer disk accretion rate, and $L_{\rm E}$ the Eddington luminosity.  The typical energy scale
shown in Figure~\ref{fig:spectra} is then $\sim 0.1$--$10\,{\rm eV}$ if the binary separation is $\sim 100 R_{\rm g}$, $\eta\sim 0.1$, and $\dot{M}$ and $M$ are not too far from $\dot{M}_{\rm E}$ and $10^8 {\rm M_{\odot}}$ respectively.  

% optical depth of the stream
We can also estimate the typical optical depth in the gas streams based on their surface density in our simulations. In code units, we find this varies within the range $\sim 0.1$--$0.5\Sigma_0$).  If we assume that electron scattering opacity dominates, the implied optical depth is
\beq
\tau_{\rm T} \simeq (92{\rm -}460)\left(\frac{\eta}{0.1}\right)^{-1}\left(\frac{\dot{M}}{\dot{M}_{\rm E}}\right)\left(\frac{a}{10^2\,R_{\rm g}}\right)^{-1/2}\,.
\label{eq:tau_sca}
\enq
Thus, for $\eta \sim 0.1$ and $\dot{M}\lesssim \dot{M}_{\rm E}$, we 
expect Thomson optical depths $\tau_{\rm T}\sim 100$ when the binary separation is about $100 R_{\rm g}$.  Adopting Kramer's law for absorption opacity, we have
\begin{eqnarray}
\tau_{\rm a} \!\!&=& \!\!3.2\times 10^{22}\,\Sigma^2 H^{-1}\, T_{\rm eff}^{-7/2} \\ \nonumber
\!\!&\simeq &\!\! (2.5 {\rm -} 65)\times 10^{-4}  
\left(\frac{\eta}{0.1}\right)^{-9/8}
\left(\frac{\dot{M}}{\dot{M}_{\rm E}}\right)^{9/8}\!\! \\ \nonumber
\!\!&\times&\!\! \left(\frac{M}{10^8 {\rm M_{\odot}}}\right)^{-15/8}\!\!
\left(\frac{a}{10^2 R_{\rm g}}\right)^{13/8}  \,.
\label{eq:tau_abs}
\end{eqnarray}
Here we adopt $\Sigma=2\rho H$ and $H\equiv c_s/\Omega(r)=0.1 (r/a)^{3/2}\,a$ with $r=1.5a$ 
for the location of the stream.  To obtain this number, we estimated the effective temperature using Equation~\ref{eq:teff_shock}; it is a factor of a few greater than $T_0$. Together with 
$\tau_{\rm T}$, we find the effective absorption optical depth is 
\begin{eqnarray}
\tau_* \!\!&\simeq&\!\! \sqrt{\tau_{\rm T}\,\tau_{\rm a}} \\ \nonumber
\!\!&\simeq&\!\! (0.15-1.73) 
\left(\frac{\eta}{0.1}\right)^{-17/16} \!\!
\left(\frac{\dot{M}}{\dot{M}_{\rm E}}\right)^{17/16}\!\! \\ \nonumber
\!\!&\times&\!\! \left(\frac{M}{10^8 {\rm M_{\odot}}}\right)^{-15/16}\!\!
\left(\frac{a}{10^2 R_{\rm g}}\right)^{9/16}  \,.
\label{eq:tau_eff}
\end{eqnarray}
Again taking $\eta \sim 0.1$ and $\dot{M}$ close to $\dot{M}_{\rm E}$, 
we find the effective opacity is roughly order unity.  However, the scale height of our simulation ($H/R \sim 0.1$) is likely an overestimate of the scale height in a realistic disk in these circumstances, so that the absorption optical depth, which is $\propto H^{-1/2}$, might be rather larger.  If so, the spectrum radiated even by the streams would be reasonably well thermalized.

%cooling time
The temperature at the stream surface should closely track the shock 
heating rate in the interior of the stream because the cooling time 
should be at worst comparable to the disk dynamical time, and likely 
shorter.  Even using the simulation scale height, we find:
\begin{eqnarray}
t_{\rm cool}\Omega 
\!\!&\simeq&\!\! H\Omega\tau/c \\\nonumber
\!\!&\simeq &\!\! (0.64{\rm -} 0.33) 
\left(\frac{\eta}{0.1}\right)^{-1}\!\!
\left(\frac{\dot{M}}{\dot{M}_{\rm E}}\right)\!\!
\left(\frac{a}{10^2 R_{\rm g}}\right)^{-1} \,.
\label{eq:tcool}
\end{eqnarray}
In fact, the short cooling time gives some justification to our assumption of an isothermal equation of state in treating the shocks.

We close this discussion with one last remark.  In ordinary accretion disks, matter traversing a factor of several in radius must lose significant energy to do so; that raises the question of what happens to the energy the streams don't radiate. The answer is that when these shocks join the binary, they do so by striking the outer edge of one of the mini-disks surrounding each black hole.   As \citet{Roedigetal2014} showed, there is a very strong shock where this happens, in which the Compton cooling time is very short.   Thus, the energy not lost by slow thermal emission is instead lost rapidly by Compton cooling at the edge of a mini-disk. In a fashion attractive to observations, it would also be modulated with a frequency of order that of the binary. However, it plays no role in the thermal spectrum because the characteristic energy of the Compton-scattered photons is $\sim 100$~keV.

\acknowledgments
JS was supported in part by the National Science Foundation under grant PHY-1144374, "A Max-Planck/Princeton
Research Center for Plasma Physics" and grant PHY-0821899, "Center for Magnetic 
Self-Organization". JHK also acknowledges support from the NSF, but through grant AST-1516299.

%%%%%%%%%%%%%%%%%%%%%%%%%%%%%%%%%%%%%%%%%%%%%%%%%%

%%%%%%%%%%%%%%%%%%%% REFERENCES %%%%%%%%%%%%%%%%%%

% The best way to enter references is to use BibTeX:

\bibliographystyle{apj}
\bibliography{cbd} % if your bibtex file is called example.bib

\end{document}